\documentclass[twocolumn]{aastex7}

\definecolor{mygreen}{RGB}{0, 185, 118}

\newcommand{\as}[1]{{\color{mygreen}{#1}}}
\newcommand{\cm}[1]{{\color{red}{#1}}}

\begin{document}

\title{Understanding the complex morphology of a CME II: \\ how pre-eruptive conditions shape CME evolution}

\author[orcid=0000-0001-5400-2800]{Abril Sahade} 
\affiliation{Heliophysics Science Division, NASA Goddard Space Flight Center, 8800 Greenbelt Rd., Greenbelt, MD 20770, USA}
\email[show]{abril.sahade@nasa.gov}

\author[orcid=0009-0009-9799-979X]{Cecilia Mac Cormack}
\affiliation{Heliophysics Science Division, NASA Goddard Space Flight Center, 8800 Greenbelt Rd., Greenbelt, MD 20770, USA}
\affiliation{The Catholic University of America, Washington, DC 20064, USA}
\email[show]{cecilia.maccormack@nasa.gov}  

\author[orcid=0000-0002-8164-5948]{Angelos Vourlidas} 
\affiliation{The Johns Hopkins University Applied Physics Laboratory, Laurel MD 20723, USA.}
\email{Angelos.Vourlidas@jhuapl.edu}

\author[orcid=0000-0003-0565-4890]{Teresa Nieves-Chinchilla}
\affiliation{Heliophysics Science Division, NASA Goddard Space Flight Center, 8800 Greenbelt Rd., Greenbelt, MD 20770, USA}
\email{teresa.nieves@nasa.gov}

\author[orcid=0000-0001-6590-3479]{Cooper Downs} 
\affiliation{Predictive Science Inc., San Diego, CA 92121, USA}
\email{cdowns@predsci.com}

\author[0000-0002-5163-5837]{Clementina Sasso} 
\affiliation{INAF – Capodimonte Astronomical Observatory, Salita Moiariello 16, 80131 Naples, Italy}
\email{clementina.sasso@inaf.it}

\author[orcid=0000-0002-6975-5642]{Judith T. Karpen}
\affiliation{Heliophysics Science Division, NASA Goddard Space Flight Center, 8800 Greenbelt Rd., Greenbelt, MD 20770, USA}
\email{judith.t.karpen@nasa.gov}

\begin{abstract}

The morphology and heliospheric impact of coronal mass ejections (CMEs) are strongly shaped by their pre-eruptive magnetic configuration and surrounding coronal environment, yet these influences remain difficult to constrain observationally. We analyze a complex CME that erupted on 2024 October 26 using multi-viewpoint remote-sensing observations and in situ measurements. Using the physics-based CORHEL-CME magnetohydrodynamic model, we test multiple physically plausible realizations of the pre-eruptive magnetic flux rope (MFR) and background magnetic field, using agreement with the observed evolution as a constraint on the CMEs initial state. We find that modest changes in MFR footpoint location and force balance lead to substantially different coronal trajectories, enabling rapid discrimination among candidate initial states. While several configurations reproduce the CMEs large-scale propagation, realistic small-scale morphology is achieved only when a near-dated background magnetic field is employed. The resulting simulation reproduces key morphologies observed from three viewpoints without fine tuning, indicating that the inferred pre-eruptive configuration represents a robust, global solution and provides a physically consistent interpretation of their magnetic origin. Comparison with in situ shock detections highlights the role of CME–solar wind interactions in shaping heliospheric signatures, though shock arrival times remain uncertain at the $11\,$hr level. These results demonstrate that data-informed, physics-based modeling can meaningfully constrain CME pre-eruptive conditions and bridge remote and in situ observations, while emphasizing the need for timely magnetic field measurements to improve predictive capability.
\end{abstract}


\section{Introduction} \label{sec:intro}

In recent decades, our understanding of solar eruptive events has evolved significantly, mainly due to the advent of dedicated solar observation missions \citep{howard_2023}. Establishing connections between extreme ultraviolet (EUV) observations of flares and prominence eruptions, white-light observations of coronal mass ejections (CMEs), and in situ measurements of magnetic obstacles (MO) has led to the identification of the key magnetic structure responsible for energy release from the low corona into the interplanetary medium: the magnetic flux rope \citep[MFR,][and references therein]{forbes_2006,vourlidas_2014, patsourakos_2020}. 

Over the years, several analytical models have been developed to characterize the evolutionary stages of an MFR. From a remote sensing perspective, various models aim to match the density enhancements in coronagraph images to determine its global morphology and kinematics \citep[e.g.,][]{thernisien_2006} in the corona. Similarly, in situ measurements of magnetic field are used to model and interpret the MFR's cross-section, when a clear MO is present \citep[e.g.,][]{nieves_2016,nieves_2018,nieves_2019,nieves_2023, weiss_2024} further out in the inner heliosphere. 
However, establishing a connection between MFR properties derived from remote sensing and in situ measurements is not straightforward. In addition to the inherent assumptions and uncertainties in each model and the limitations of the measurements, the CME evolves between the corona and the inner heliosphere, potentially altering the MFR properties through distortion and erosion. Thus, analytical models must account for these dynamic processes to accurately describe the MFR evolution, \citep[see e.g.,][and references therein]{nieves_2023b,alhaddad_2025}.

In parallel, numerical simulations offer a physics-based approach to study the CME evolution. Currently, the models most widely used for CME heliospheric propagation are the magnetohydrodynamic (MHD) solvers ENLIL \citep{enlil} and EUHFORIA \citep{euhforia}. These codes are computationally inexpensive, making them well-suited for ensemble studies and forecasting applications. This computational efficiency comes at the cost of domain limitations (starting above $20\,R_\odot$) and reduced physical realism of CME dynamics and complexity \citep[see e.g.,][]{maharana_2024}. A major limitation is that most of the CME deformation and reorientation occurs below $\sim15\,R_\odot$, where magnetic forces dominate. Missing this early-stage complexity adds considerable uncertainties in predicting the CME arrival time (averaging $13\,$hours) and geoeffectiveness \citep{vourlidas_2021,kay_2024}.

Currently, only three modeling frameworks simulate the full evolution of CMEs from the solar surface to Earth: CORHEL-CME \citep{linker_2024}, SWMF \citep{swmf}, and COCONUT+EUHFORIA \citep{linan_2025}. Runs for the first two can be requested via the Community Coordinated Modeling Center (\href{https://ccmc.gsfc.nasa.gov}{CCMC}) at NASA’s Goddard Space Flight Center through a user-friendly web interface. These physics-based thermodynamic MHD models allow comprehensive investigations of the MFR evolution and its interactions with background solar wind, while they also provide synthetic imagery that can be directly compared with observations. A better understanding of these interactions is essential for advancing our physical knowledge of CME evolution in the inner heliosphere and is crucial for assessing the space-weather impacts of CMEs at various locations \citep{vourlidas_2019,winslow_2022}.

In a previous study, \citet{maccormack_2025b} demonstrated the ability of CORHEL-CME to perform simulations of real events. We were able to connect unusual white-light structures to their magnetic origins, underscoring the value of combining high-resolution observations with physics-based modeling. 

In this paper, we analyze an event that occurred on 2024 October 26 (hereafter the October event). The October event was observed remotely from three vantage points, whereas its broad shock impacted four widely separated heliospheric missions. We investigate the CME with the objective of connecting and physically understanding its evolution from the low corona to the inner heliosphere. Using the publicly available CORHEL-CME model, we perform Sun-to-1au simulations constrained by multi-viewpoint remote observations and evaluate their consistency with in situ shock detections at four spacecraft.
The October event presents uncertainties in its pre-eruptive magnetic configuration, requiring exploration of multiple physically plausible realizations of the magnetic flux rope and background field. We assess which configurations can self-consistently reproduce the observed coronal morphology and global propagation, and examine how differences in the initial magnetic system influence the CMEs large-scale evolution. While heliospheric shock signatures provide only partial constraints on internal flux rope geometry, their multi-point detections offer valuable information on the CMEs global energetics and expansion \citep{palmerio_2025}.
Through this approach, we evaluate the ability of physics-based, data-informed modeling to provide a coherent Sun-to-heliosphere interpretation of a complex CME, and to identify the limitations imposed by uncertainties in the background magnetic field and ambient solar wind. 

Section~\ref{sec:obs} summarizes the available data for the October event. In Section~\ref{sec:analysis} we describe the methods and techniques used in this study to model the event. Section~\ref{sec:results} presents the results of the simulations and their comparison with the data. Discussion and conclusions are provided in Sections~\ref{sec:disc} and \ref{sec:conc}.

\section{Observations}\label{sec:obs}

\begin{figure*}[]
\epsscale{1.2}
\begin{interactive}{animation}{JHV_filament.mp4}
\plotone{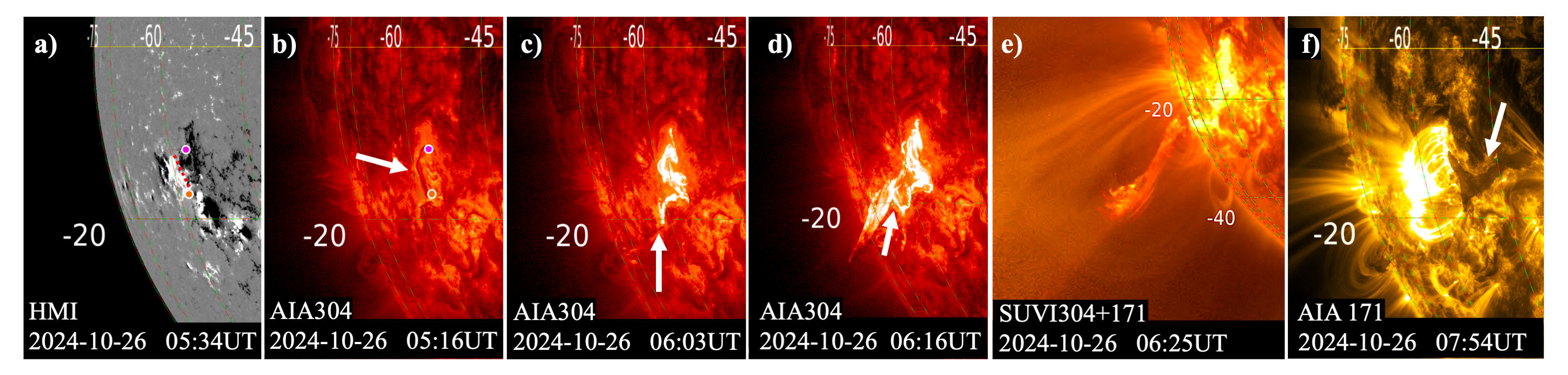}
\end{interactive}
\caption{Pre-eruptive configuration and early eruption of the October event. a) HMI magnetogram of AR 13873. The red dashed line marks the PIL. The magenta (orange) dot indicates the northern (southern) footpoint. b) pre-eruption filament observed in AIA 304 at 05:16$\,$UT (white arrow). Footpoints are marked as in panel a). c) First activation at 06:03$\,$UT, with flare ribbon signatures. The southern extension of the ribbons is indicated by the white arrow. d) Eruptive filament at 06:16$\,$UT, highlighted by the white arrow.  e) Early eruption at 06:25$\,$UT observed by SUVI, shown as a composite of the 304 and 171 \AA{} channels to show the core of the eruption and the ambient corona. f) Post-eruptive arcade observed in AIA 171 at 07:54$\,$UT. White arrow indicates the location of the dimming associated with the northern footpoint. An animated version of this figure shows the evolution of the eruption in AIA 304, SUVI 304, SUVI 131 and SUVI 171 images from 05:00 UT to 09:00. \label{fig:Filament}}
\end{figure*}

\begin{figure*}[]
\epsscale{1.15}
\plotone{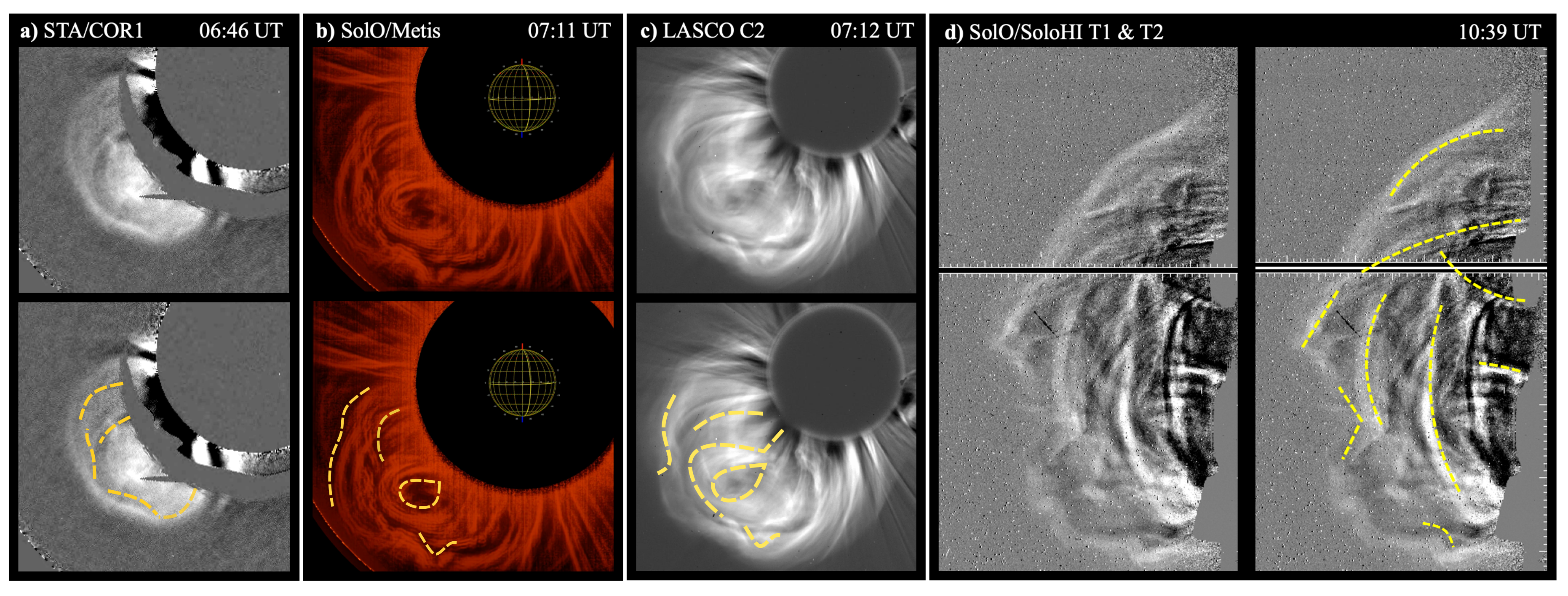}
\caption{Early CME evolution. a) COR1 base-difference images  at 06:46$\,$UT.  Dashed curves denote features mentioned in the main text (Sec.~\ref{sec:obs}). b) Polarized brightness Metis observations  at 07:11$\,$UT processed with the wavelet-optimized whitening (WOW) filter that enhances the edges. c) Similar to panel a) for LASCO C2 at 07:12$\,$UT. d)  SoloHI inner FOV (tiles 1 and 2) running-difference images. 
\label{fig:CME}}
\end{figure*}

On 2024 October 26, an eruptive M9.5 flare in active region NOAA 13873 activated at around 06:00 UT. The eruption was observed simultaneously by the Atmospheric Imaging Assembly \citep[AIA;][]{lemen_2012} and the \textit{Helioseismic and Magnetic Imager} \citep[HMI;][]{scherrer_2012}, onboard the Solar Dynamics Observatory \citep[SDO;][]{pesnell_2012}; the Solar Ultraviolet Imagers \citep[SUVI;][]{suvi_2022} on board the Geostationary Operational Environmental Satellite (GOES); and the Extreme Ultraviolet Imager \citep[EUVI;][]{wuelser_2004} of the Sun-Earth Connection Coronal and Heliospheric Investigation suite \citep[SECCHI][]{howard_2008} suite onboard the Solar TErrestrial RElations Observatory Ahead \citep[hereafter STA;][]{kaiser_2008}.
We use the AIA observations to follow the on-disk evolution of the eruption because of their high spatial and temporal resolution, and the extended field of view of the SUVI images to track the eruption off limb.

The pre-eruptive filament overlies the north-south polarity inversion line (PIL) of AR 13873, indicated with a red dashed line over the HMI magnetogram in Figure~\ref{fig:Filament}(a). It extends from latitude $-11^\circ$ to $-17^\circ$ along $\sim -56^\circ$ longitude, as shown in the 304$\,$\AA{} image (Figure~\ref{fig:Filament}(b)). Both panels show the footpoint locations, identified by the magenta (negative polarity) and orange (positive polarity) dots. The location of the southern footpoint is particularly interesting, as it lies at the closest approach between ARs 13873 and 13872, which may play a role in the eruption. At $\sim$05:58$\,$UT, the first signs of heating appear on the southern portion of the filament. By 06:03$\,$UT most, if not all, of the filament and the surrounding area has been heated to MK temperatures, as evidenced by the 131$\,$\AA{} movie (included in the animated version of Fig.~\ref{fig:Filament}) with hot loops appearing northwards of the filament. From 06:03 to 06:23$\,$UT, the heating expands eastward from the southern footpoint, indicating reconnection over a longer PIL segment, or possibly along a different PIL. This interpretation is further supported by the progression of the southern ribbons. Figure~\ref{fig:Filament}(c) and (d) show brightening in the AIA 304$\,$\AA{} channel, with the drift of the flare ribbons indicated by the white arrow. The 304$\,$\AA{} brightenings extend about $20^\circ$ east of the original filament location. From about 06:08 to 06:25$\,$UT, a rising arcade/MFR system becomes increasingly clearer, while 304$\,$\AA{} material is drawn into it and thermalized (Fig.~\ref{fig:Filament}(d,e)). Dimmings associated with the northern portion of the eruption are evident for another $13^\circ$ west (Fig.~\ref{fig:Filament}(f)). East of the eruption, dimmings cannot be discerned due to foreshortening and interference from the post-eruption arcade. Note that the final post-eruption arcade extends along the north-south axis for about $15^\circ$. The full evolution of the event, including post-eruption loops and dimmings, is shown in the animated version of Figure~\ref{fig:Filament}, which includes the AIA 304$\,$\AA{}, SUVI 304$\,$\AA{}, SUVI 131$\,$\AA{} and SUVI 171$\,$\AA{} filters from 05:00 to 09:00 UT.

\begin{figure*}[]
\epsscale{1.17}
\plotone{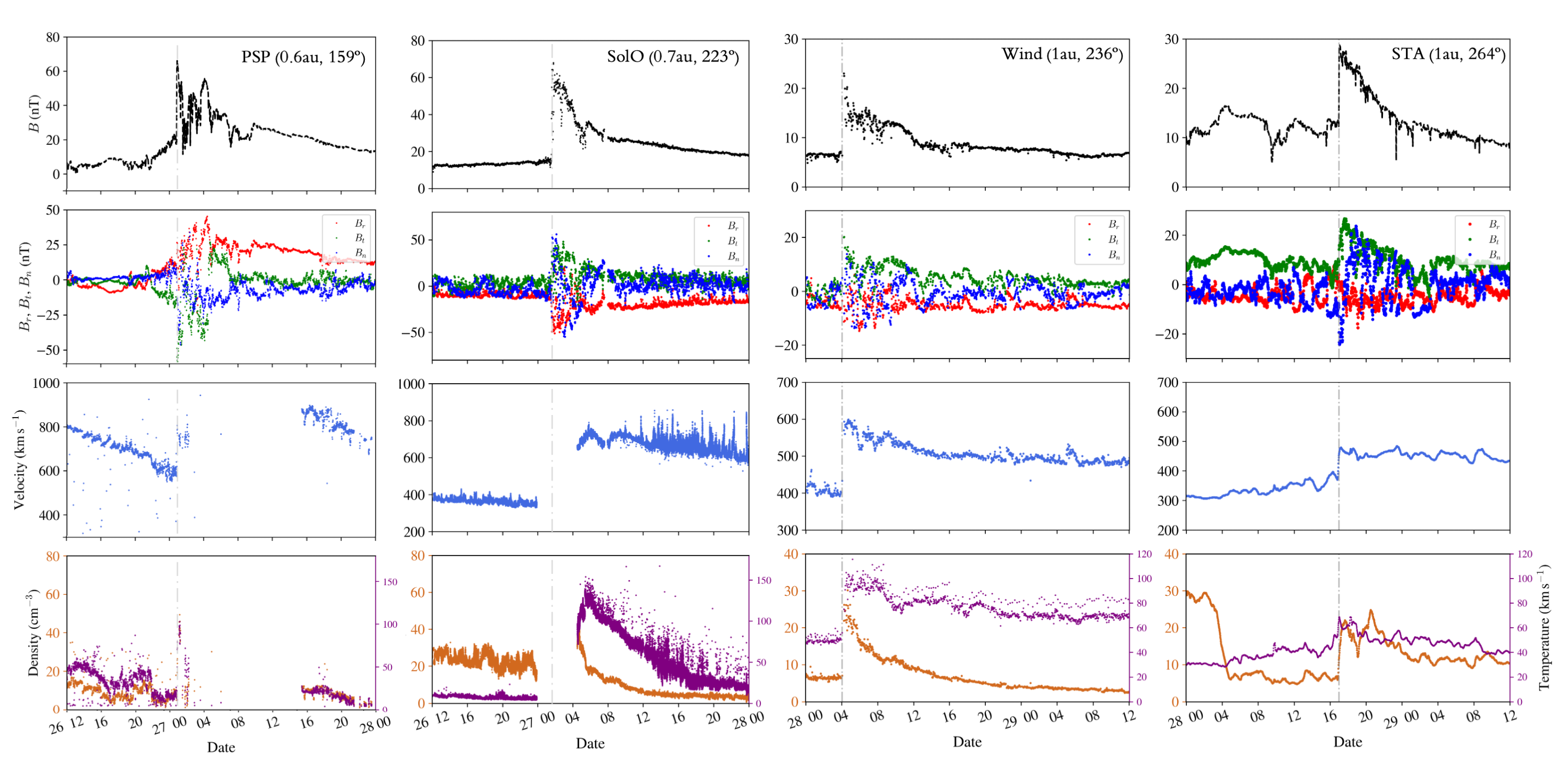}
\caption{Solar wind data during October 26-31 provided by PSP, SolO, Wind and STA, respectively. From top to bottom we show the total magnetic field [nT], magnetic field component (RTN) [nT], solar wind velocity [km$/$s] and density [cm$^{-3}$]. Dashed gray vertical lines indicate the CME shock arrival. \label{fig:IS}}
\end{figure*}

The resulting CME was observed by the Large Angle and Spectrometric Coronagraphs \citep[LASCO;][]{brueckner_1995} C2 and C3 on board the Solar and Heliospheric Observatory \citep[SOHO;][]{domingo_1995}, the COR1 and COR2 coronagraphs of the SECCHI suite onboard STA, the Metis coronagraph \citep{antonucci_2020} onboard the Solar Orbiter mission \citep[SolO;][]{muller_2020}, and the Solar Orbiter Heliospheric Imager \citep[SoloHI;][]{howard_2020}. STA and SolO were located at $27.7^{\circ}$ and $-14.0^{\circ}$ relative to the Sun-Earth line, respectively.  We also tracked the evolution of the CME from the low corona (using AIA and EUVI  171 \AA~) to the heliospheric imagers, using the Solarsoft \texttt{rtclowdwidget} routine that fixes footpoints of the Graduated Cylindrical Shell (GCS) model \citep{thernisien_2006, thernisien_2009} at $1\,R_\odot$ and allows the CME axis to be tilted to follow the deflection \citep[see, e.g.,][]{sahade_2023}. The apex of the CME is deflected about $(40\pm10)^\circ$ to the south, with the legs located at ($180^\circ$,$-12^\circ$). The average speed is $(1400\pm100)\,$km s$^{-1}$. Uncertainties are estimated by varying the GCS model parameters about the best-fit solution until the model no longer provides an acceptable visual match to the observations; the corresponding parameter ranges are reported as the errors.

The CME was first detected in COR1 at around 06:21 UT, while the hot MFR is still visible in the AIA FOV (Fig.~\ref{fig:Filament}(d)). The initial three-part structure of the CME starts to get distorted in the southern part of the front around 06:51$\,$UT. This distortion remains notable throughout the COR2 FOV, as well as a northern shock evolving with the CME.

Similar shapes are seen from the LASCO C2 and C3 images starting around 06:36 UT. A clearer definition of the inner structure with an MFR signature towards the southeast is visible. A density enhancement in the CME core creates a ``question mark'' shape (see dashed lines of Fig.~\ref{fig:CME}c), and a distorted or corrugated front is created by the interaction with the surrounding magnetic field in the south. 

At 06:51 UT, the CME first appears in the Metis FOV, within the polarized visible light channel. Although it had a similar perspective to LASCO ($\sim 14^\circ$ of angular separation), these polarized brightness observations are processed with a wavelet-optimised whitening \citep[WOW;][]{WOW} filter that enhance the edges and are better to identify the intricate inner structure of the CME (Figure~\ref{fig:CME}). Well-defined concentric ovals indicate the location of the MFR in the center of the CME propagating toward the southeast. Moreover, the corrugated front is observed ahead of the southern part while a less distorted front on the east side can be interpreted as part of the CME shock and corresponds to the northern portion of the shock observed by COR1.

SoloHI observations capture the later evolution of the CME (Figure~\ref{fig:CME}d). SoloHI observations have higher spatial resolution and sensitivity than STA/HI-1 and hence they are crucial for our work. The MFR signature is detected in tile 1 of the detector and a less distorted shock is seen in tile 2.

Figure~\ref{fig:CME} presents snapshots of the CME as seen from the different spacecraft. The highlighted features (dashed yellow lines over the images) will be used for comparison with the CORHEL-CME  simulations discussed in the next Section. 

The wide shock of the October event was detected in situ by Parker Solar Probe \citep[PSP;][]{fox_2016} at ($0.6\,$au, $159^\circ$, $3^\circ$), SolO at ($0.7\,$au, $223^\circ$, $8^\circ$), Wind at ($1\,$au, $236^\circ$, $5^\circ$) and STA at ($1\,$au, $264^\circ$, $2^\circ$). The coordinates of the spacecraft correspond to the Carrington longitude and latitude at the time of the eruption (06:00 UT). 
Figure~\ref{fig:IS} shows the detected total magnetic field and its components, and the solar wind velocity, density and temperature for PSP, SolO, Wind and STA.
None of the spacecraft show signatures of a coherent magnetic field rotation typically associated with a flux rope passage, most likely due to the southward deflection of the CME. As a result, the in-situ observations are limited to the detection of the shock and do not provide direct constraints on the internal magnetic structure of the MFR. However, they remain valuable for providing benchmarks for the simulations and help constrain the global energetics on the CME propagation through the heliosphere.

At Wind and STA, we found a sharp increase of the magnetic field, speed, proton density and temperature associated with the interplanetary shock wave arrival. Unfortunately, at PSP and SolO, gaps in the plasma moments (speed, density, and temperature) datasets prevent us from fully characterizing the shock but the arrival is clearly identified by the sharp increase in the magnetic field magnitude. 
The time of arrival at PSP was on 2024 October 27 00:49$\,$UT, at SolO on 2024 October 27 at 01:33$\,$UT, at Wind on 2024 October 28 at 04:14$\,$UT, and at STA on 2024 October 28 at 17:00$\,$UT.
Both Wind and STA present signatures of a previous arrival that is not entangled with the October 26 event, but preconditions the solar wind background that the new arrival had encountered. 

To understand and connect all the observations we use the CORHEL-CME model to reproduce the full evolution of the October event. As we mentioned before, this eruption presents significant complexity in its early evolution due to its complex source region and because the EUV instruments onboard SDO, GOES, and STA observed the eruption from the west side, which caused foreshortening on the east side.  This raises the question of how to determine the initial magnetic configuration: Is the filament’s observational signature alone sufficient to reconstruct the location of the pre-eruptive MFR, where then the interaction with the magnetic environment will play a more crucial role in shaping the eruption? Or should we consider an initial scenario where free energy is present on a different portion of the PIL, following the path of the observed flare ribbons, to better reproduce the eruption? 
In this study, we aim to explore how variation in the initial MFR configuration can provide the best match to the later CME morphology observed in the corona (observed concentric ovals, southward distorted front and question mark shape), as well as in situ shock observations. The methods used to achieve these goals are outlined in the following section.

\section{Methods and analysis}\label{sec:analysis}

\begin{figure*}[]
\epsscale{1.17}
\plotone{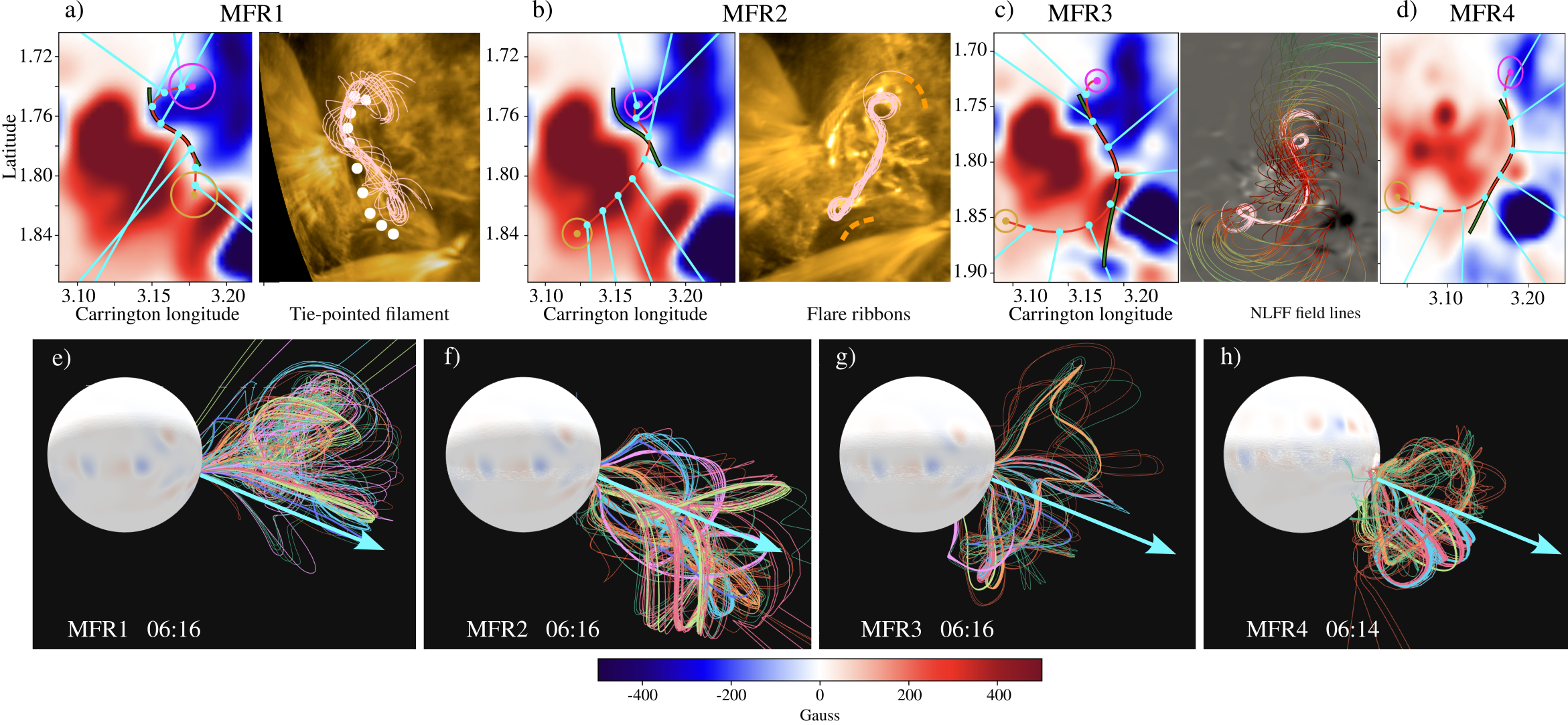}
\caption{ Modeling of the MFRs. For each MFR, top left panels show the magnetogram and the segment of the PIL (solid green line) over which the MFR was built. The orange (magenta) dots indicate the positive (negative) footpoint, and the red line connecting them indicates the path of the MFR. The Lorentz force is calculated over the cyan dots and the cyan lines indicate the force direction. Right top panels show, for each case, some field lines of the initial MFR (pink lines) with the context of the selection: the tie-pointed position of the pre-eruptive filament (white dots) for MFR1 (a), the flare ribbons location (dashed orange lines) for MFR2 (b), and the magnetic field (color-coded field lines) derived from the NLFF model for MFR3 (c), and MFR4 (d). Bottom panels (e-h) show the magnetic field lines of the MFRs in the early evolution, and the cyan arrow indicates the apex direction of the GCS model from the observations. \label{fig:MFRs}}
\end{figure*}

CORHEL-CME \citep{linker_2024}\footnote{\url{https://ccmc.gsfc.nasa.gov/models/CORHEL-CME~1/}} allows users to perform simulations of CMEs within a realistic solar corona and solar wind environment. A key feature of the framework is its use of the Regularized Biot–Savart Law \citep[RBSL;][]{titov_2018} to generate MFRs that can follow complex filament channels and PILs in active regions. Users initiate the process by selecting the CME source region on a magnetogram and specifying the relevant segment of the PIL as well as the positive and negative footpoints anchoring the MFR. From these inputs, CORHEL-CME builds an MFR using the RBSL model, which can then be customized by the user in terms of its physical properties. The CORHEL-CME interface offers zero-beta MHD simulations (i.e. only magnetic forces are considered) for rapid evaluation of the eruptive potential of the MFR configuration. Once the initial parameters are selected, users can request a full-physics, thermodynamic MHD simulation that evolves the CME from the low corona ($1\,R_\odot$) out to $1\,$au, embedded within a self-consistently reconstructed background solar wind. 

The early evolution of the October event poses a challenge for the initial modeling of the MFR, particularly in determining the portion of filament channel that erupts. To deal with this challenge we explore different ways of selecting the location of the MFR footpoints based on the available observations.
Following the methodology of \citet{maccormack_2025b}, we build a first MFR (MFR1) based on the 3D coordinates of the pre-eruptive filament determined with the tie-pointing technique applied to AIA and EUVI images \citep[see, e.g.,][]{inhester_2006}. 
As mentioned in Section \ref{sec:obs}, the flare ribbons indicate an eastward drift of the positive-polarity footpoint of the eruptive system. Therefore we use the position of the flare ribbons to build a second MFR (MFR2). 
Due to the difficulty of reproducing the observed complexity only by tracking plasma parcels, we also perform a non-linear force-free (NLFF) reconstruction of the magnetic field. We used the \href{https://github.com/RobertJaro/NF2}{NF2} package version 0.3.0 \citep{jarolim_2023} to extrapolate SHARP 12094 at 05:00 UT on 2024 October 26. This reconstruction yields a new location of the footpoints, from which a third MFR (MFR3) was built. The western portion of the extrapolated field lines is spatially consistent with the observed filament location, providing qualitative support for the reconstructed magnetic field. We note that the NLFF extrapolation is used here as a proxy for the magnetic configuration, and its results are considered in conjunction with those obtained from the observationally constrained methods (MFR1–MFR2).

The complexity of the source region presents a modeling challenge, but the remote-sensing observations, combined with later multi-point in-situ measurements, provide strong constraints on the model results. The event offers a valuable opportunity to evaluate the MHD model's sensitivity to initial conditions, and to determine its ability to reproduce the observations from the solar surface to 1 au.

\subsection{MFR modeling}

The base magnetogram used to set the bottom boundary conditions of the simulations is the synoptic daily HMI map of 2024 October 26. The daily synoptic maps update the region between longitude ($-60^\circ$,$60^\circ$) at the observed time. At this time, the source region is quite close to the -60 cutoff in the daily synoptic data.

As described in Section \ref{sec:obs}, the southern end of the eruptive filament appears to activate a more distant section of the filament channel, leading to the eruption of a larger portion of the magnetic system. Based on the 3D position of the filament and the locations of the flare ribbons, we build two MFRs with different footpoint configurations, hereafter MFR1 and MFR2. With the NLFF reconstruction of the source region we build a third MFR, hereafter MFR3. The main difference between the MFRs is the choice of the eruptive portion of the PIL and the location of the positive footpoint. Figure~\ref{fig:MFRs} shows the selected segment of the PIL (green solid line) and the footpoint locations in the magnetogram, for each case. Note that the northern (negative) footpoint position needs to be adjusted to ensure flux balance in the MFR. Figure~\ref{fig:MFRs}(a,b) show MFR1 and MFR2, respectively, for which the same portion of the PIL is selected to match the triangulated position of the filament (white dots in panel a). However, the MFR2 positive footpoint is displaced to match the position of the flare ribbons at 06:02 UT (orange dashed lines of panel b). This time was chosen as an early state of the eruption, before a clear rise of the system. For MFR3, a longer segment of the PIL is selected to better follow the magnetic field shape extrapolated with the NLFF model (Figure~\ref{fig:MFRs}c). To achieve an unstable (eruptive) configuration, the axial current within the RBSL model MFR must exceed the equilibrium threshold for restraint by the strapping field above the PIL. In CORHEL-CME this is typically expressed as a fractional current above this threshold. In our case, zero-beta simulations were run to test the eruptivity, using similar absolute axial currents for all cases. The bottom panels of Figure~\ref{fig:MFRs} show a snapshot of the evolution of the MFRs $\sim 15\,$ minutes after the eruption onset. The morphological differences among the three erupted MFR configurations are obvious. The simulations show that the case with the positive footpoint located to the west (MFR1) led to a more radial eruption, whereas those with the positive footpoint to the east (MFR2 and MFR3) produce a southward-directed eruption that agrees more closely with the observations. On the other hand, the anchoring of the negative footpoint also modifies the MFR evolution, in the sense that MFR3 exhibits both northward and southward components, like a mix between MFR1 and MFR2. 

However, in these simulations the CME interacts with magnetic structures farther east that are not yet updated in the daily synoptic map at this time. Therefore, we created a new base magnetogram that combines the synoptic maps of Carrington rotation 2289 (2024-09-19 to 2024-10-16) and 2290 (2024-10-16 to 2024-11-13) with the central portion of the daily magnetogram ($-43^\circ$,$42^\circ$) to obtain a different realization of the measurements in this portion of the map. We then model MFR1, MFR2, and MFR3 with the new base magnetogram in the zero-beta simulations. They exhibit similar dynamics as in the previous background in terms of propagation direction, although there are some differences in evolution. In particular, MFR2 deflects less to the south in the new background, making MFR3 with the new background the only one matching the observed direction of propagation. Therefore, along with the results of MFR1, MFR2, and MFR3 over the HMI daily background, we present the results for MFR3 in the new background as MFR4 (Fig.~\ref{fig:MFRs}(d)). 

The full simulation parameters and outputs are available at the CCMC website: \href{https://ccmc.gsfc.nasa.gov/ror/results/viewrun.php?runnumber=Abril_Sahade_051425_SH_1}{MFR1}, \href{https://ccmc.gsfc.nasa.gov/ror/results/viewrun.php?runnumber=Abril_Sahade_052725_SH_1}{MFR2}, \href{https://ccmc.gsfc.nasa.gov/ror/results/viewrun.php?runnumber=Abril_Sahade_110625_SH_3}{MFR3}, \href{https://ccmc.gsfc.nasa.gov/ror/results/viewrun.php?runnumber=Abril_Sahade_111925_SH_1}{MFR4}. In the next section we present the differences among the four MFR simulations and compare the results with the described data.

\section{Results}\label{sec:results}

\begin{figure*}[]
\epsscale{1.17}
\plotone{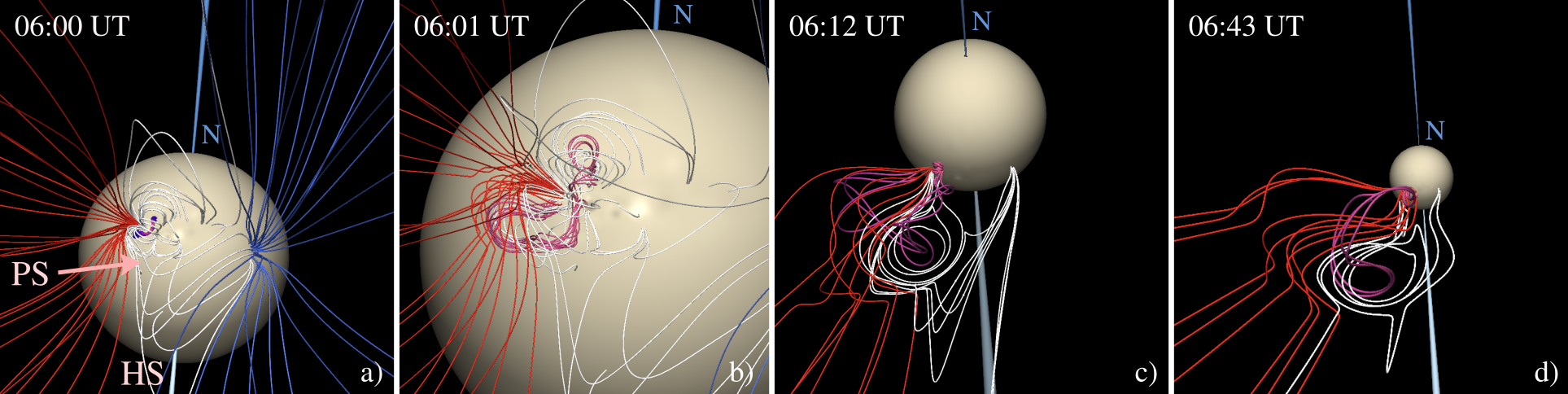}
\caption{Magnetic field line evolution of MFR4 and its environment. The purple field lines indicate the MFR axis. The open field lines of positive (negative) polarity are shown in red (blue) and the closed overlying field is drawn by the white lines. The north (N) and south pole are indicated in each panel in light blue to depict the orientation. a) the initial state of the simulation (at 06:00 UT). The overlying pseudostreamer (PS) and helmet streamer (HS) are indicated. b) 06:01 UT, at which the southern part of MFR4 started to lift and interacted with the eastern coronal hole (red open field lines). c) 06:12 UT, showing the reconnection of the closed field beneath the MFR axis and the ongoing perturbation of the coronal hole field lines. d) a later state (06:43 UT) in which the MFR4 axis is less writhed while it keeps evolving through its magnetic environment.\label{fig:MFR4}}
\end{figure*}
\subsection{Coronal evolution}
To decide which pre-eruptive model best represents the October event, we compare the CME evolution from the three vantage viewpoints of COR1, LASCO/C2, and Solar Orbiter/Metis and SoloHI. Figure~\ref{fig:MFRs}(e) shows that MFR1 does not match the observed direction of propagation of the CME (cyan arrow). Therefore MFR1 traced from its initial position does not accurately reproduce this event.
For MFR2 and MFR3 (Fig.~\ref{fig:MFRs}(f,g)), the propagated field lines agree with the three-dimensional apex locations of the GCS model obtained from the observations (indicated by the cyan arrow). However, a more detailed comparison of the magnetic field lines with the white-light observation does not show many features in common. For example, the axis orientation of the MFRs can not explain the concentric ovals and question mark shape pointed out in the observations (Fig.~\ref{fig:CME}). 

\begin{figure*}[]
\epsscale{1.17}
\begin{interactive}{js}{3d_MFR4.html.zip}
\plotone{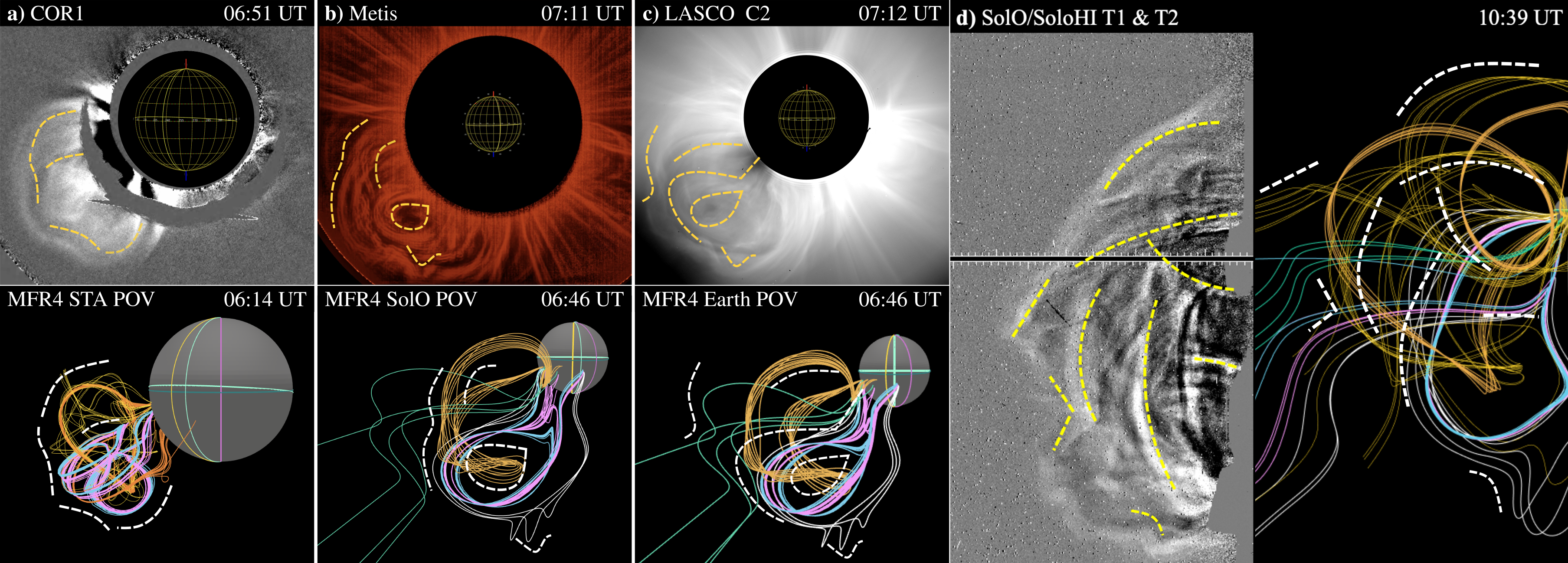}
\end{interactive}

\caption{White-light CME images and modeled field line evolution of MFR4. The top panels show the CME of October event as observed by STA COR1, SolO Metis, LASCO C2,and the tiles 1 and 2 of SoloHI. The middle panels show the field lines of MFR4 in the perspective of STA, SolO, Earth, and SolO again for a later snapshot (from left to right, respectively). For reference, the meridians and FOV of each instrument are plotted in pink for STA, yellow for SolO and green for LASCO and the solar equator. Lastly, lower panels show the white-light observations merged with the MFR field lines from the model, making more evident the match between the observed morphology and subjacent magnetic structure. Note that, although the size of the Sun is matched for the comparison between observations and simulation, the FOVs do not match perfectly due to a projection effect of the 3D rendering of the field lines, for which the eye of the camera is closer than the actual telescopes. This effect is more notable for the right panel at SoloHI, for which is not possible to pile up the images; instead some common features are highlighted by dashed lines. \label{fig:CME_MFR}}
\end{figure*}

In contrast,  we find a compelling match between the observed morphology and the shape of the magnetic field lines for MFR4 (Fig.~\ref{fig:MFRs}(h)). To understand the key features that shaped MFR4, Figure~\ref{fig:MFR4} presents snapshots of its temporal evolution. Panel~(a) shows the initial state of the simulation. Red (blue) field lines denote positive (negative) open magnetic fields originating from the coronal holes that surround the eruptive system. White field lines outline the closed overlying magnetic topology above MFR4 (marked by the purple lines). The MFR is embedded within one lobe of a pseudostreamer topology (indicated by PS), whose spine is connected to the helmet streamer (HS) bounded by the surrounding coronal holes.
Panel~(b) illustrates the onset of the eruption. The southern portion of the MFR rises first, likely due to weaker confinement by the overlying magnetic field, and begins interacting with nearby open field lines (red). As the eruption progresses, the overlying arcade expands and the northern portion of MFR4 is also lifted.
Panel~(c) helps explain several observed CME features. Flare reconnection occurs beneath the rising MFR, in the eastern lobe of the pseudostreamer. This reconnection simultaneously adds flux to the MFR, producing projected concentric rings around the MFR axis, consistent with the rings observed from the Metis perspective (see white field lines).
The MFR fast evolution does not give the helmet streamer field lines time to relax, thus forming a corrugated front on top of the MFR. Lastly, the expansion of the MFR also distorts the helmet streamer field lines producing a complex shock feature to the north of the MFR. Panel (d) shows how these features remain throughout the evolution, including the overlapping of field lines to explain the ``question mark'' feature from LASCO POV. Moreover, this initial configuration makes the MFR axis evolve in a direction that is consistent with the cavity observed in the white-light images of Metis, LASCO, and STA, and allows a portion of MFR4 to evolve to the east, forming a shock in the direction of PSP. 

Figure~\ref{fig:CME_MFR} directly compares the CME evolution from the different white-light imagers' perspectives and the corresponding projections of the magnetic field lines of the modeled MFR4. Figure~\ref{fig:CME_MFR} presents the October CME as observed by COR1 (a),  Metis (b), LASCO/C2 (c), and  SoloHI (tiles 1 and 2, d) with the mentioned features. The bottom panels display the MFR4 field lines from the perspectives of STA, SolO, Earth, and SolO, again at the times when MFR4 matches the height of the GCS modeling for the observations. Orange field lines correspond to the axis of MFR4. The cyan and pink field lines show an outer envelope of the MFR that reconnected with the overlying arcade near the western negative footpoint. The white field lines correspond to the helmet streamer arcade and the green field lines indicate the open field lines to the east. For reference, the meridians and FOVs of each instrument are indicated on the solar surface: pink for STA, yellow for SolO, and green for LASCO, along with the equator. The dashed lines of each panel represent the common features outlined in the observations and projected field lines, highlighting the correspondence between the observed CME morphology and the underlying magnetic structure. An animated version of Figure~\ref{fig:CME_MFR} provides a 3D view of panels b and c, that allows the reader to hover, zoom and change the perspective of the presented field lines. 

As the eruption evolves, the eastern part of the MFR interacts with the helmet streamer (as explained above and shown in Figure~\ref{fig:Filament}(f)), producing the shock features observed to the north in the coronagraph images. On the other hand, the extra flux generated by the pink and cyan field lines wrapped around the main axis of the MFR leads to density enhancements observed in the southern portion of the CME; the MFR axis produces the cavity feature in the center. As the POV changes from each spacecraft these field line projections look distinctive: from the COR1 POV it appears as a double hump CME (Fig.~\ref{fig:CME_MFR}(a)); Metis detects the oval-shaped features (Fig.~\ref{fig:CME_MFR}(b)); and from LASCO/C2 they look more like a question mark  (Fig.~\ref{fig:CME_MFR}(c)). 
Later, the MFR field lines are still useful for comparing the observations of SoloHI and recognizing their magnetic origin. Some of the matching fronts and features are highlighted in the rightmost panel of Fig.~\ref{fig:CME_MFR}.

\subsection{Heliospheric evolution}
\begin{figure*}[]
\epsscale{1.05}
\plotone{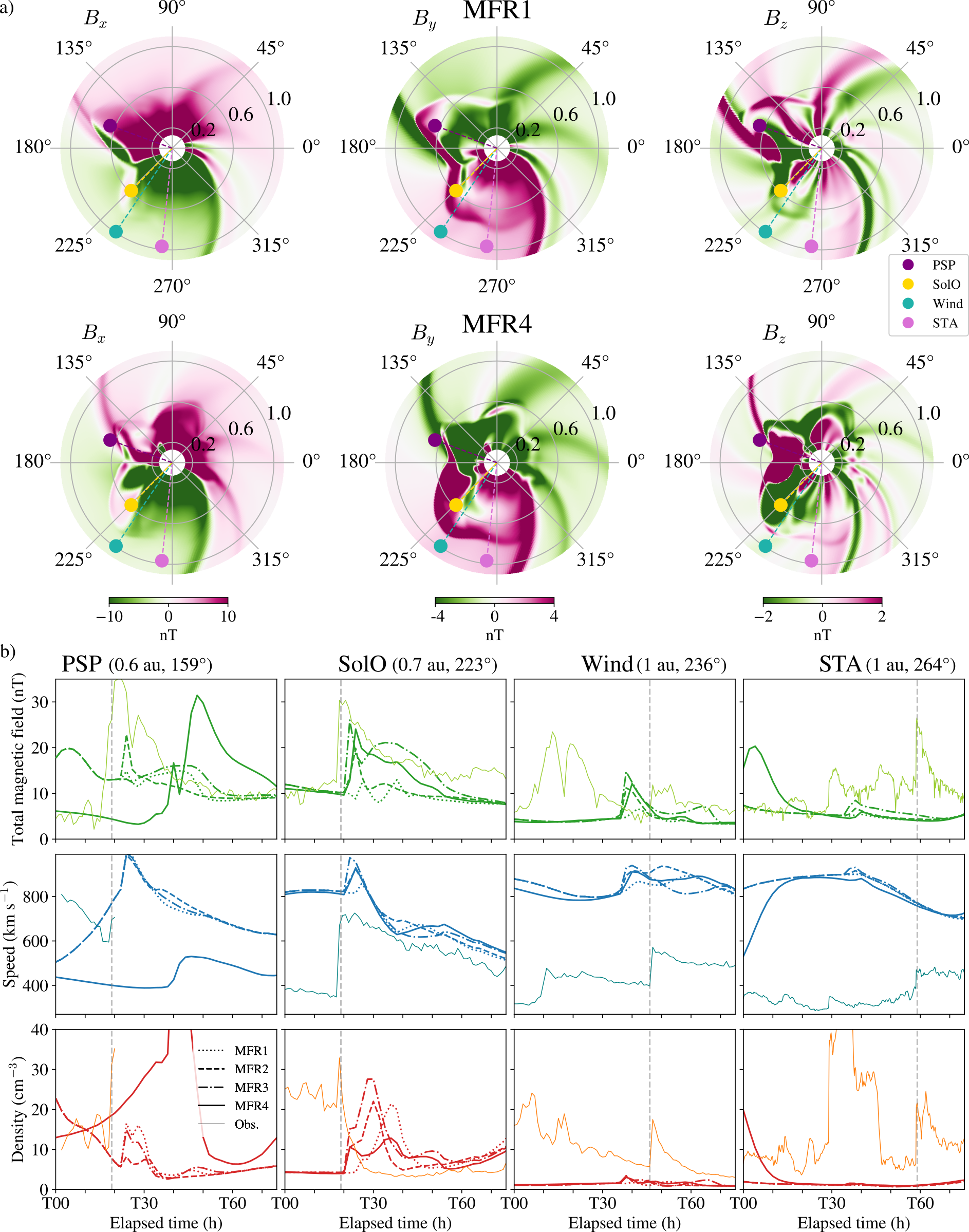}
\caption{a) Slices of the magnetic field components at latitude $6^\circ$ at 34 hours into the MFR1 and MFR4 simulations. The dots indicate the heliospheric position of STA (pink), Wind (teal), SolO (yellow) and PSP (purple). b) Comparison of simulation parameters and observed in-situ data at different heliospheric locations. From left to right: PSP, SolO, Wind, and STA measurements of total magnetic field (top panels), solar wind speed (middle panels), and solar wind density (bottom panels). The x-axis expresses the time elapsed from the initiation of the eruption in hours. The gray vertical dashed lines indicate the time of arrival in the observed data. \label{fig:MFRs_B}}
\end{figure*}

Despite the differences in the coronal evolution of the simulated MFRs, their heliospheric evolution show some commonalities, in particular in the speed of the shock, which is largely driven by the free energy (which was selected to be similar in all the simulations) and the background solar wind speed. The CME evolved on both sides of the heliospheric current sheet (HCS) located nearly perpendicular to the solar equator at Carrington longitude $\sim 210^\circ$ for the background of MFR1, MFR2, and MFR3, and at $\sim 180^\circ$ for the background of MFR4. The HCS generated a high density region that compressed the CME front. Moreover, slow and fast wind streams also modified the CME speed along its front, contributing to its distortion. Figure~\ref{fig:MFRs_B}(a) shows longitudinal slices of the magnetic field components of MFR1 and MFR4 34 hours after the initiation of the eruption (i.e., 2024 October 27 at 16:00 UT). The slices were taken at latitude $6^\circ$, since all spacecraft were located near that plane on that date. The CME front is most clearly outlined in the middle panel ($B_y$ component). The positions of the spacecraft are displayed with colored dots, for context. The evolution of MFR1, MFR2, and MFR3 is similar. Some major differences in the fronts of MFR1 and MFR4 are evident due to the different streamer location in each background (see, for example, the CME fronts in the $90^\circ$ to $180^\circ$ quadrant at $B_y$). 

As shown before (Figure~\ref{fig:IS}) all of the spacecraft had been impacted by the CME shock but do not all have a clear MO signature. Figure\ref{fig:MFRs_B}(b) compares the measured and simulated total magnetic field, speed, and density for each spacecraft. The simulated magnetic field magnitude of the four simulations is scaled by a factor of 2 to facilitate comparison with in-situ measurements. Such underestimation of the magnetic field strength at 1 au is a known limitation of CORHEL-CME framework, with previous studies reporting discrepancies of a factor of $\sim$1.5–2 \citep{linker_2016,torok_2018}, likely due to model simplifications and underestimation of the photospheric magnetic field in the input magnetograms.
In this study, the scaling factor is further constrained by the requirement that the background solar wind magnetic field (before and after the CME passage) matches the observed values. A factor of 2 provides consistent agreement across all four spacecraft locations (two at 0.6 au and two at 1 au), supporting the robustness of this choice. The times are expressed in hours since the initiation of the eruption (2024 October 26, at 06:00 UT). 
The measured time of arrival is indicated by dashed gray lines. To assign an uncertainty to the time of arrival derived from the simulations, we consider both the fluctuations around the measurement point (i.e., variations with respect to neighboring grid cells) and the discrete output times of the simulation. The resulting uncertainty is $\pm1$~h for all models.

\begin{deluxetable}{lcllll}
\tablewidth{0pt}
\tablecaption{Time of arrival in hours after the eruption for observations and modeled flux ropes (MFR1–MFR4). All values have an uncertainty of $\pm1$~h. \label{tab:toa}}
\tablehead{
\colhead{S/C} & \colhead{Obs} & \colhead{MFR1} & \colhead{MFR2} & \colhead{MFR3} & \colhead{MFR4}
}
\startdata
PSP   & 19 & 23 \cm{(+ 4)} & 23 \cm{(+ 4)} & 23 \cm{(+ 4)} & 38 \cm{(+19)} \\
SolO  & 19 & 29 \cm{(+10)}& 21 \cm{(+ 2)}& 21 \cm{(+ 2)} & 21 \cm{(+  2)} \\
Wind  & 46 & 39 \as{( - 7)} & 37 \as{( - 9)} & 37 \as{( - 9)} & 37  \as{( -  9)} \\
STA   & 59 & 38 \as{( -21)}& 36 \as{( -22)}& 35 \as{( -23)}& 39 \as{( -20)}\\
\enddata
\end{deluxetable}

Table~\ref{tab:toa} summarizes the time of arrival, expressed in hours after the eruption, for the observations and the four modeled flux ropes (MFR1–MFR4) at PSP, SolO, Wind, and STA. From these values, we derive the average absolute error of each model relative to the observations, yielding errors of $\pm10.5$~h, $\pm9.5$~h, $\pm10.0$~h, and $\pm12.5$~h for MFR1, MFR2, MFR3, and MFR4, respectively.

The arrival time errors exhibit an asymmetry, with delayed arrivals at PSP and SolO and early arrivals at Wind and STA. Several factors likely contribute simultaneously. First, the source region is located close to the cutoff in the daily synoptic data, resulting in reduced accuracy of the eastern portion of the coronal magnetic field. This may affect the reliability of the reconstructed coronal and heliospheric magnetic field, particularly toward the east where PSP and SolO are located, and consequently impact the CME propagation and predicted arrival times (see Figure~\ref{fig:MFRs_B}).
Second, a previous CME modified the solar wind conditions at Wind and STA, an effect not included in the simulations. Third, the use of a steady-state solar wind introduces an intrinsic limitation, as it cannot capture the temporal evolution of the heliosphere and its large-scale structures.

Because the spacecraft are distributed over a wide longitudinal range ($\sim100^\circ$) and at different heliocentric distances (0.6–1 au), these factors do not affect all locations equally, making it difficult to attribute the asymmetry to a single cause. The fact that the model is not systematically early or late at all locations suggests that the global CME propagation speed is reasonably captured, while local discrepancies are primarily driven by inaccuracies in the background solar wind and CME–environment interactions.
Finally, no clear magnetic obstacle signatures are observed at any spacecraft, preventing a quantitative comparison of flux rope duration. The in-situ validation is therefore limited to shock-related quantities.

The largest difference in the arrival times among the MFRs occur at PSP, reflecting the sensitivity of the CME propagation to the location of large-scale structures such as the HCS. In particular, MFR4 arrives later due to compression by the high-density, slow solar wind associated with the HCS. Despite this delay, the magnetic field profile of MFR4 shows better agreement with the observations than the other configurations. This may indicate that MFR4 more accurately captures a CME–HCS interaction occuring at PSP, although this interpretation remains uncertain due to gaps in the plasma data during that interval.
At SolO, MFR2, MFR3 and MFR4 arrive at the same time but exhibit distinct magnetic field profiles. Again, MFR4 provides the closest match to the observations. The post-shock solar wind speed and density are also well reproduced. At Wind and STA, all MFRs arrive at similar times, with a relatively weak impact at STA. Interestingly, STA observed a stronger magnetic enhancement than Wind despite experiencing a more glancing encounter. In the MFR4 simulation, an increase in magnetic field and density around T10 is associated with the HCS, suggesting that a CME–HCS interaction may contribute to the observed enhancement. However, this interpretation remains uncertain due to the presence of a preceding CME, not included in the model, that complicates signatures in the observations. 

Although MFR4 provides the best agreement with the remote-sensing observations and reproduces the in-situ magnetic field profiles most accurately, it does not yield the smallest arrival time errors. This highlights that whereas the early evolution of the CME is well constrained by observations, the heliospheric comparison of shock arrival times is more sensitive to uncertainties in the background solar wind.

\section{Discussion}\label{sec:disc}

Solar eruptive events comprise a chain of intricate processes that have the potential to greatly affect our technology. Efforts to understand the complexity of these events during their propagation through the heliosphere, and leverage the existing capabilities to improve our knowledge about CMEs, are crucial. Bridging the gap between remote-sensing and in situ observations remains a key challenge in achieving a continuous, self-consistent description of a CME evolution.

In this work, we modeled the full evolution of the 2024 October 26 CME using the publicly available CORHEL-CME model. By combining multi-viewpoint white-light observations with multi-point in situ shock detections, we constructed and evaluated end-to-end simulations of a complex event, from its pre-eruptive magnetic configuration to its large-scale heliospheric expansion. To our knowledge, this represents one of the first comprehensive Sun-to-1 au applications of this model in a fully data-informed, multi-viewpoint context, enabling a better understanding of the complex evolution that a CME undergoes as it propagates.

Because the eruptive portion of the filament channel for this complex event was not easy to determine, we implemented different methods to build the MFR that best represents the pre-eruptive system using magnetic and EUV information. Rather than relying on extensive parameter tuning, our approach emphasizes physically motivated exploration of plausible initial states, using agreement with observed CME evolution, as well as filament location, flare ribbon evolution, and a NLFF model, to constrain the pre-eruptive magnetic system. We found important differences in the direction of propagation of the MFR depending on the footpoint location, due to its magnetic structure and its interaction with the ambient field. 

We followed the full evolution of four initial configurations to evaluate their differences in coronal and heliospheric evolution. The best-performing case (MFR4), which incorporates an updated background magnetic field, represents a robust global solution rather than a narrowly optimized configuration. This result highlights the dominant role of the pre-eruptive magnetic system and its coupling to the ambient field in shaping CME morphology during the magnetically driven early phase of evolution. Figure~\ref{fig:CME_MFR} shows some examples of this best match that emphasize the magnetic structure underlying the density enhancements we observe in the white-light images. This kind of data-informed modeling can help us to understand how close or far we are from the standard flux rope morphology and to improve the interpretation of the observations in terms of the CME magnetic complexity.

During its interplanetary evolution the October event produced a large shock that impacted four heliospheric missions, which were dispersed across $\sim100^\circ$ of longitude and different heliospheric distances (PSP and SolO at $\sim 0.6\,$au, and Wind and STA at $1\,$au). While no MFR signatures were detected by the spacecraft, the arrival of the shock at each probe provides valuable data for validating CME propagation models and constraining their initial free energy.
Our results had on average a $\sim 11\,$h error in arrival time at each spacecraft, comparable to that currently estimated for forecasting models \citep{kay_2024}. However, it is important to consider that, for two spacecraft, a previous event conditioned the 
heliosphere and the model can not reproduce that. Moreover, at PSP we did not have updated magnetic field information and our results showed the importance of the interactions with the HCS, which determined the shock profile and arrival time. 

These results underscore the importance of having timely and updated photospheric magnetic field measurements to properly capture the large-scale magnetic environment in which CMEs evolve. Future assets, such as Vigil, will be key in providing upstream information on magnetic maps and solar wind parameters, improving the realism of CME initial conditions for both state-of-the-art simulations and forecasting. 

More generally, this event highlights the complementary nature and inherent limitations of remote-sensing and in-situ observations. Remote observations provide strong constraints on the early evolution and morphology of CMEs, but suffer from projection effects and incomplete viewpoint coverage. In contrast, in-situ measurements offer precise local diagnostics of the plasma and magnetic field, yet sample only a single trajectory through the structure and, in this case, do not capture the magnetic flux rope itself.
In this context, physics-based, data-informed MHD modeling can provide a coherent, end-to-end interpretation of a complex CME, successfully linking coronal magnetic structure to large-scale heliospheric impact. The combination of multi-viewpoint observations enables meaningful discrimination among plausible pre-eruptive configurations and constrains the global energetics of the eruption. At the same time, the present results illustrate that such models remain sensitive to uncertainties in the background solar wind and boundary conditions. Therefore, the interpretation of complex events relies on an iterative approach, in which observations constrain the models, and models, in turn, help interpret incomplete observations.

\section{Conclusions}\label{sec:conc}

The most important conclusions drawn from this work are:
\begin{itemize}
    \item The pre-eruptive magnetic configuration strongly governs CME evolution, with modest differences in flux rope connectivity and force balance producing qualitatively different trajectories and morphologies.
    \item Exploration of plausible initial states shows that only a subset can reproduce the observed evolution, allowing other configurations to be ruled out and identifying robust global solutions rather than finely tuned cases.
    \item The magnetically dominated early phase of CME evolution sets CME morphology and internal structures. The interaction of the erupting flux rope with the surrounding coronal magnetic environment is revealed by the remote sensing observations and, further out, constrained by the multi-point heliospheric in situ data. 
\end{itemize}

Overall, this study demonstrates the strength of combining multi-viewpoint remote imaging, multi-point in situ measurements, and physics-based models to follow and interpret the coronal and heliospheric evolution of CMEs. Such approaches advance our physical understanding of solar eruptions, and represent essential steps toward improving the predictive capability of next-generation space-weather models.

\begin{acknowledgments}

We thank the anonymous reviewer for helpful and constructive comments that improved the clarity of this work. A.S. is supported by an appointment to the NASA Postdoctoral Program at the NASA Goddard Space Flight Center, administered by Oak Ridge Associated Universities under contract with NASA. C.M. is supported by the NASA Heliophysics Division, Solar Orbiter Collaboration Office under DPR NNG09EK11I. A.V. is supported by NASA grant 80NSSC22K1028 and 80NSSC22K0970. T.N.-C. thanks the support of the Solar Orbiter mission. C.D. is supported by the NASA Living With a Star Strategic Capabilities program: NASA grant 80NSSC22K0893. Simulation results have been provided by the Community Coordinated Modeling Center (CCMC) at Goddard Space Flight Center through their publicly available simulation services (https://ccmc.gsfc.nasa.gov). The CORHEL-CME Model was developed by Predictive Science Inc. 
Solar Orbiter is a mission of international cooperation between the European Space Agency (ESA) and the National Aeronautics and Space Administration (NASA), operated by ESA. The Solar Orbiter Heliospheric Imager (SoloHI) instrument was designed, built, and is now operated by the US Naval Research Laboratory with the support of the NASA Heliophysics Division, Solar Orbiter Collaboration Office under DPR NNG09EK11I. Animations from the observations were produced with JHelioviewer \citep{JHelioviewer}. Figure~\ref{fig:MFR4} was produced with HelioSpace, developed and adapted by Timothy Hall to read CORHEL-CME data. 
This research was supported by the International Space Science Institute (ISSI) in Bern, through ISSI International Team project 'Understanding the Onset of Solar Eruptions' (ISSI Team project \#24-606).

\end{acknowledgments}

\begin{contribution}

AS conceived the study, performed the CORHEL–CME simulations and associated analyses, and drafted the manuscript. CM contributed to the study preparation and development of the research concept, processed the images, and contributed to the analysis and comparison between observations and model outputs, as well as to the discussion and editing of the manuscript. AV contributed to major revisions of the manuscript and to the interpretation of the EUV and white-light observations. TNC contributed to the interpretation of the in situ observations. CD provided the modified magnetograms for the different background configurations. CS provided the Metis data. CD, CS, and JK contributed to manuscript preparation.

\end{contribution}

\facilities{SolO/Metis, SDO/AIA, STEREO-A/EUVI, SOHO/LASCO-C2, STEREO/COR1 and SolO/SoloHI data}

\bibliography{Oct_26}{}
\bibliographystyle{aasjournalv7}

\end{document}